\documentstyle[11pt,ngxo]{article}
\begin{document}
\title{\large\bf Science Driven arguments for a 10~sq.meter,
1~arcsecond X-ray Telescope}

\author{Martin Elvis and Giuseppina Fabbiano}
\author{Harvard-Smithsonian Center for Astrophysics,
Cambridge Mass USA}


\begin{abstract}

X-ray astronomy needs to set bold, science driven goals for the
next decade. Only with defined science goals can we know what to
work on, and a funding agency appreciate the need for significant
technology developments.  To be a forefront science the scale of
advance must be 2 decades of sensitivity per decade of time. To
be stable to new discoveries these should be general, discovery
space, goals.

A detailed consideration of science goals leads us to propose
that a mirror collecting area of 10~sq.meters with arcsecond
resolution, good field of view ($>$10~arcmin), and with high
spectral resolution spectroscopy ($R$=1000-10,000) defines the
proper goal. This is about 100 times AXAF, or 30 times XMM. This
workshop has shown that this goal is only a reasonable stretch
from existing concepts, and may be insufficiently bold.

An investment of $\sim$\$ 10M/year for 5 years in X-ray optics
technologies, comparable to NASA's investment in ASTRO-E or a
SMEX, is needed, and would pay off hugely more than any small
X-ray mission.

\end{abstract}

%

\vspace{-0.3in}

\section{Long Term Goals for X-ray Astronomy}

Any big undertaking, such as X-ray astronomy
\footnote{
We discuss only the standard 0.1-10~keV band of X-ray astronomy,
in which grazing incidence optics work efficiently.}
surely is, must set long range goals. With clear long-term
science goals in place we can see which developments are
essential, and which are mere sidelines - amusing, but dead
ends. Daniel Goldin, the NASA Administrator, urged astronomers
(San Antonio AAS meeting, January 1996) to make decade length
plans, even if the plan changes in a few years time. This is what
we do here.

The goals we shall describe are deliberately ambitious.  We
propose that X-ray astronomers should aim to reach sensitivities
100 times beyond AXAF, while retaining high angular resolution
and achieving high dispersion spectroscopy (Table
\ref{goal})
\footnote{
These numbers describe quite well Hale's Mt.Wilson 100-inch
telescope (Osterbrock 1995), so we are suggesting that
X-ray astronomy try to equal the state of optical astronomy in
1917.}
. This does not mean that the very next mission we design should
necessarily have all these capabilities. Nor does it rule out
smaller missions with different goals. It does mean that the next
major mission we design should at least be a deliberate and
significant step toward these capabilities. To some readers
these goals may seem hopelessly idealistic. As a result of this
workshop, it seems instead that they are only a few factors of 2
away from reality.

In section 2 we outline several areas of wide astrophysical
importance to which X-ray astronomy can make crucial
measurements, given a 10~sq.meter,1~arcsec. telescope ; in
section 3 the current state of X-ray astronomy is reviewed and
general `discovery space' arguments are introduced; ; in section
4 we use these arguments to derive the 10~sq.meter, 1~arcsec,
$>$10~arcmin field of view telescope goals; in section 5 we
discuss instrumentation goals including the need for sensitivity
in the 0.1-0.5~keV band; in section 6 we summarize our assessment
of the workshop against these goals; and in section 7 we open the
discussion on how to make this telescope a reality.

\begin{table}
\vspace{-0.3in}
\begin{center}
\caption{The 10 sq. meter X-ray Telescope}
\label{goal}
\vspace{0.05in}
\begin{tabular}{|ll|}
\hline
Large Area: & {\bf 10 sq. meters}\\
Good Angular Resolution: & {\bf 1~arcsec}\\
Wide Field Imaging: & {\bf $>$10 arcmin}\\
Good Spectral Resolution: & {\bf 1000$-$10,000}\\
\hline
\end{tabular}
\end{center}
\vspace{-0.4in}
\end{table}

\section{Science Drivers}

Where should X-ray astronomy be in the first decades of the 21st
century? X-ray astronomy brings a unique potential for
understanding the distant universe to astrophysics. Hot plasmas,
matter under extreme conditions (density, temperature, pressure),
and under intense gravitational fields, can only be directly
observed in the X-ray band.  These observations have implications
for cosmology, the life cycle of matter, and relativity.

However, today X-ray astronomy can barely detect ordinary
galaxies at the distance of Coma, clusters of galaxies at
$z\sim$0.5, and quasars at z=4 - all regimes where optical
infrared and radio astronomy are flourishing.  The X-ray missions
to be launched in the late `90s will help.  With AXAF and XMM, in
typical 10$^4$s exposures, X-ray astronomers will be able to get
a CCD spectrum with $R$=E/$\Delta$E$\sim$10-80 (with $\sim$1000
counts) for any source brighter than 10$\mu$Crab (the RASS
limit), about 10$^5$ sources in the sky. This sounds powerful,
but if optical telescopes were similarly limited to the brightest
10$^5$ objects they would stop at V=12, missing Pluto, the Crab
pulsar, LMC cepheids (and hence the distance scale), and quasars
altogether
\footnote{Contrast this with
optical astronomy where 10$^8$ objects are accessible (to B=21,
Jones et al., 1991).}.
What is X-ray astronomy missing by not being able to sample a
larger universe of objects? Even more limiting is the fact that
AXAF and XMM have no more area for grating spectroscopy than the
ROSAT PSPC had for imaging.  Consequently, they will take
$>$10$^6$s to get an $R$=100 spectrum (which is several
times worse than even the most routine optical spectrum) of
a 10$\mu$Crab X-ray source. What physics is being missed by
lacking basic spectral diagnostics?

We have compiled a short list, covering topics ranging from
cosmology to the solar system, of the obvious gaps that proposed
missions cannot fill, but where a 10 sq.meter 1~arcsec. telescope
will make crucial contributions. These topics are not of interest
only to high energy astronomers, but are of wide-ranging
astrophysical relevance. With telescopes of the right sensitivity
(Table \ref{10sqm}), X-ray astronomy will play a key role in unraveling
the fundamental mysteries of the Universe.

\medskip
\noindent\underline{\em Geometry of the Universe (H$_0$,
$q_0$), and Large Scale Flows:}
X-ray astronomy provides unique methods of probing the geometry
of the Universe. The `baryon catastrophe' and the abundances of
the hot intracluster medium (ICM) in high redshift clusters of
galaxies are already indirect constraints on the cosmological
model. The Sunyaev-Zeld'ovich effect is the most straightforward
of the physics-based techniques for measuring $H_0$ and $q_0$. It
combines X-ray and radio astronomy and can be used to large
redshifts.

\small
\begin{table}[t]
\begin{center}
\caption{Capabilities of a 10 sq. meter X-ray Telescope}
\label{10sqm}
\vspace{0.05in}
\label{capabilities}
\begin{tabular}{|lllcrl|}
\hline
Target& no. sources&flux(cgs)&$\mu Crab$& counts     & Type of spectra$^b$\\
      & in sky$^a$ &         &          & in 10$^4$s &  	       \\
\hline
Bright sources& 3,000 &2.10$^{-12}$ &100 &10$^6$& $R$=1000 abs'n  \\
              &       &           &    &      & $30\times 30$CCD  \\
\hline
ROSAT Sky Survey& 100,000& 2.10$^{-13}$&10&10$^5$&R=10,000 em.\\
(RASS) limit&          &           &      &  & R=1000 abs.\\
{\bf Baseline Flux, $f$}&  &           &      &  & $10\times 10$CCD\\
\hline
ROSAT medium& 4 million&2.10$^{-14}$& 1&10$^4$ & $R$=1000 \\
 survey     &          &           &     & &  {\bf $3\times 3$CCD }\\
\hline
ROSAT Deep Survey&30 million &2.10$^{-15}$&0.1& 1000& CCD  \\
(1st NLXG)    &           &          &   &    &            \\
\hline
AXAF Deep& 100 million &2.10$^{-16}$&0.01 & 100 & colors \\
Survey        &             &          &     &    &      \\
\hline
Fully Resolved& 100 million? &2.10$^{-17}$&0.001 & 10 & detection \\
CXRB?         &             &          &     &    &      \\
\hline
\end{tabular}

a. from ROSAT extragalactic logN-logS (0.3-2.4 keV
band). Multiply by $\sim$3 for 2-10 keV (Inoue et al., 1996);
b. $R$=E/$\Delta$E=$\lambda/\Delta\lambda$.
\end{center}
\end{table}
\normalsize

Purely X-ray observations can also be used to measure H$_0$ and
$q_0$ in a direct way. This method employs the absorption lines
produced by the ICM in the spectrum of a background quasar, plus
the cluster emission measure (Krolik \& Raymond, 1988). High
angular resolution is needed to reduce the background in the
quasar spectrum due to the cluster, and large collecting area to
build a spectrum. Thousands of Abell clusters have quasars for
which a $R=1000$ spectrum can be obtained with a 10~m$^2$ X-ray
telescope in 10$^4$s (1$\mu$Crab) so large numbers of accurate
H$_0$ measurements can be made quickly. Their dispersion will map
large scale motions. $q_0$ can be measured with z$>$0.2 clusters
in a total of 10$^6$s.

\medskip
\noindent\underline{\em Deep Surveys:}
The origin of the cosmic X-ray background (CXRB) has been the
`Holy Grail' of X-ray astronomy for over 30 years. It is time not
only to solve this mystery definitively, but to go beyond it, to
understand the nature of the CXRB creating sources.  Two thirds
of the CXRB is now resolved into point sources, mostly AGN. What
is the remaining 1/3? Probably something new. The increase in the
logN-logS curve at faint fluxes (0.1$\mu$Crab, Hasinger et al
1993) seems to be made of `Narrow Line X-ray Galaxies' (NLXG,
Jones et al., 1996) of unknown nature. A 10m$^2$ X-ray telescope
can obtain $R=1000$ spectra of the brightest of this supposed new
source population, which may be galaxies undergoing an early
burst of star formation. In 10$^4$s a 10m$^2$ X-ray telescope
with 1$^{\prime\prime}$ resolution and a 30 arcmin dia. field of
view can detect 1~nanoCrab (100 times fainter than ROSAT), and
one 10$^5$s exposure will gather good $R=10$ spectra of 200-600
of these sources (Table \ref{10sqm}), allowing investigation of
their nature. The $\sim$250 brightest sources will give X-ray
redshifts directly from the Fe-K line, if they have Raymond-Smith
like spectra, or are heavily obscured AGN. In 10$^5$s ROSAT Deep
Survey sources will give $R$=1000 spectra.

\medskip
\noindent\underline{\em The Intergalactic Medium:} The IGM left
over from galaxy formation was barely constrained in its
properties until recently. The IGM has now been detected
by HST and HUT (Jakobsen et al 1994, Davidsen et al., 1995);
Lyman-$\alpha$ forest clouds have been recognized to fill large
amounts of intergalactic space; and these same clouds are found
to be `polluted' by early metal production.

However the UV Helium features are hidden by Galactic absorption
at low z. Only X-ray spectra allow the investigation of the
evolution of the IGM down to the present.  X-ray absorption
features toward low z AGN will measure abundances in the local
IGM, and map out its ionization state as a function of
environment (Aldcroft et al., 1994).  A 10sq.m X-ray telescope
can sample thousands of lines of sight at $R=1000$.  Moreover, if
deep surveys resolve 99\% of the CXRB into discrete sources then
the $\Omega$ allowed in a hot diffuse IGM can be well constrained
from limits on its bremsstrahlung emission (Barcons \& Fabian
1992).

\medskip
\noindent\underline{\em Quasar Environments at z$\geq$4:}
At high redshift there is dramatic optical evidence (the
`alignment effect') that quasars and radio galaxies lie in a high
density medium which probably has X-ray temperatures. Strong,
highly polarized, Lyman-$\alpha$ emitting gas with knotty complex
shapes is seen aligned with the radio structures and may mark
sites of jet-induced star formation.

A 10~sq.meter X-ray telescope can image this hot medium down to
$L_X=3\times 10^{43}erg~s^{-1}$ in 10$^4$s at z=3
(0.01$\mu$Crab). The Lyman-$\alpha$ emission extends over only a
few arcseconds (for any z$>$2 1~arcsec.$\sim$20~kpc), and central
quasars are bright, so 1~arcsecond imaging is essential.
Studying this medium in absorption against the luminous ($L_X\sim
10^{47}erg~s^{-1}$) central quasar will give abundances and
ionization state (and distances).  $R=1000$ X-ray spectra of many
quasars will be obtainable to z$>$4.

\medskip
\noindent\underline{\em Protogalaxies:}
X-ray spectroscopy adds new and unique capabilities to current
optical/UV studies of non-luminous material seen toward
quasars in absorption. The strongest of these absorbers are the
damped Lyman-$\alpha$ systems which are likely to be the
long-sought protogalaxies (Wolfe 1995). While common at high z
they die out toward the present, presumably forming the galaxies
we now see around us.

X-ray K-edges give virtually ionization independent measures of
total column density and so accurate abundances (unlike the
saturated optical/UV lines of individual ions) and ionization
states (Madejski et al., 1995). X-ray absorption lines probe hot
ISM states inaccessible to lower energy observations, and can be
studied at 1000 times smaller column density than edges, opening
up the lower column density `metal line systems' that lie around
more developed galaxies (Steidel et al., 1995). The quasars
involved are faint (10$\mu$Crab) and so 10~sq.meters of
X-ray collecting area are essential.

\medskip
\noindent\underline{\em Masses of quasar black holes:}
ASCA has shown good evidence that broad Fe-K lines exist in AGN
X-ray spectra.  These surely originate close the the event
horizon of the central black hole, probably from the inner edge
of an accretion disk. These lines are broadened by Doppler shifts
and by the General Relativistic redshift escaping from the hole's
potential. These distortions allow us to study matter in strong
gravity, and to measure the `mass function' of quasar black holes
through cosmic time.  To determine a line profile needs $\sim$20
bins, each determined to $\sim$5\%. A typical Fe-K line carries
$<$5\% of an AGN X-ray flux, so 10$^5$ counts are needed, and
hence to extend this measurement to high redshift
($\leq$10$\mu$Crab) an area of 10~sq.meters is essential. To
employ reverberation mapping on bright AGN needs the same area.

\medskip
\noindent\underline{\em Spectroscopy \& Plasma diagnostics:}
The application of spectroscopy and atomic physics to astronomy
transformed it into astro\underline{physics}. X-ray astronomy
will likewise be transformed by high dispersion spectroscopy. New
plasma diagnostic tools over a wide range of celestial objects -
stellar coronae, SNR, the hot ISM of galaxies, the ICM of
clusters, ionized absorbers in AGN - will become
available. Spectroscopy also yields Doppler velocities.
Table \ref{R1000} lists a few applications.

\begin{table}
\vspace{-0.3in}
\begin{center}
\caption{{\em R=1000} Spectroscopy with the 10sq. meter X-ray
Telescope}
\label{R1000}
\vspace{0.05in}
\begin{tabular}{|l|}
\hline
$\bullet$Quasar environments at z$>$3 (via absorption spectra)\\
$\bullet$Protogalaxy physics (via damped Ly-$\alpha$ absorbers)\\
$\bullet$Existence and masses of quasar black holes\\
$\bullet$Quasar black holes: strong gravity GR\\
$\bullet$Quasar accretion flow doppler mapping\\
$\bullet$Compact binary orbits, masses, environments\\
$\bullet$Stellar winds: profiles, enrichment of ISM, \\
$\bullet$Coronal tomography\\
$\bullet$Chemical evolution in Galaxies/clusters to z=1\\
$\bullet$SS433-like double-jet sources out to M51 \\
$\bullet$SN expansion velocities \& abundances to Virgo.\\
\hline
\end{tabular}
\end{center}
\vspace{-0.2in}
\end{table}

The X-ray emission line spectrum is rich in plasma diagnostic
capabilities:
\newline\underline{Electron density} diagnostics are derived from
the line ratios of forbidden ($z$) to intercombination
($x$,$y$) lines of the He-like ions.
\newline\underline{Temperature} diagnostics come from the ratio
of dielectronic satellite lines to resonance lines through the
Boltzmann factor.  The ratio ($x + y + z)/w$, where $w$ is the
resonance line, is also temperature-sensitive.
\newline
\underline{Abundance} derivations in a multitemperature plasma
require a fully consistent analysis of the temperature
distribution..

High spectral resolution ($R>1000$) is essential for these
diagnostics. Line blending is pervasive throughout the spectral
region below 2~keV where the great majority of diagnostics
lie. In solar spectra with $R=400$, most of the bins contain
contributions from half a dozen lines (McKenzie et al., 1980),
making unambiguous decomposition impossible. At $R=1000$ the line
blending eases considerably.

Little attention has been given to the potential of {\em really
high resolution} spectra ($R$=10,000) in X-ray astronomy. At this
resolution new physics becomes available, through the measurement
of thermal widths and doppler motions (Table \ref{R10000}) At
$kT=$1~keV, with ions and electrons in equipartition, a medium
mass ion such as oxygen has a thermal velocity of
$\sim$100~km/s. To measure this width requires a spectral
resolution of $\sim$30~km/s, i.e. $R$=10,000.  The inherent limit
to X-ray spectral resolution due to thermal line widths only
applies to thermal, not photoionized plasmas, and only cuts in
for $R>$10,000 and kT$>$10~keV.

\small
\begin{table}
\vspace{-0.3in}
\begin{center}
\caption{High Spectral Resolution ($R$=10,000) Spectroscopy}
\vspace{0.05in}
\label{R10000}
\vspace{0.05in}
\begin{tabular}{|p{5in}|}
\hline
{\bf Thermal width}
Ion temperatures can be determined for each ion, so testing both
whether the ions are isothermal, and whether they are in thermal
equilibrium with the electron temperatures derived from
bremsstrahlung.
\\ \hline
{\bf Doppler Motions}
are often of order 100~km/s. In binary systems, stellar winds,
spiral galaxies, and mergers of clusters of galaxies this is a
good benchmark value. Line centroids can be measured to this
level even with lower resolution spectra, but not if two
different velocity components are blended, as is likely to be the
case in many systems. SNR turbulent velocities could be measured
out to the LMC, and binary Doppler shifts (and so masses) out to
M~31.
\\ \hline
{\bf Absorption Lines}
can come from really cold plasma, as low as 10~K in star forming
molecular clouds. Narrow lines are best seen at high
resolution. Absorption line cross-sections are of order 1000
times those of absorption edges. So absorption line studies open
up a large range of absorbers to potential study: instead of
N$_H\sim 10^{21}$ to detect oxygen we need only N$_H\sim
10^{18}$.
\\ \hline
{\bf Photoionized plasmas}
create new opportunities at R=10,000. The thermal temperature of
a photoionized plasma will be lower than that of the incoming
radiation and so will have smaller, physically interesting, line
widths. The higher line density emitted by photoionized plasmas
also benefits from higher spectral resolution.
\\ \hline
\end{tabular}
\end{center}
\vspace{-0.3in}
\end{table}
\normalsize

\medskip
\noindent\underline{\em Galaxies:}
Galaxies are key objects for the study of cosmology, the life
cycle of matter, and stellar evolution. X-ray observations allow
the study of key components such as the hot ISM, evolved compact
objects, and active nuclei, which are either impossible or hard
to explore in other spectral bands. With these data we can
measure the metal abundance in the hot ISM of galaxies and the
surrounding cluster medium; when hot X-ray haloes are present we
can measure galaxy masses (and baryonic content and thus set
constraints on $\Omega$); we can study the evolution of the
gaseous component and thus learn about global evolution of
galaxies and matter life cycle; we can gain crucial understanding
of the formation and evolution of binary X-ray sources, and for
the evolution of SN and SNR; we can explore the faintest active
nuclei and understand their nature.  Galaxy ecology - the study
of the cycling of enriched materials from galaxies into their
environment - is {\em inherently} an X-ray subject. Escape
velocities from galaxies, when thermalized, are kilovolt X-ray
temperatures.

AXAF and XMM provide too few photons for these faint objects, and
XMM, with its $>$10~arcsec HPD, also lacks the resolution to
cleanly separate individual sources and galaxy components, except
in a handful of the nearest galaxies.  In Table \ref{galaxies} we
show the depth of study that a 10~sq.m.  X-ray telescope
allows. Detailed feasibilities come from Fabbiano (1989) and from
table \ref{10sqm}. Arcsecond resolution is needed to resolve
individual sources out to Virgo, the nearest place with a sample
of elliptical galaxies.

By looking at large samples of galaxies over a large lookback
time ($z>1$), we can study the evolution of global X-ray
properties. We can obtain X-ray data on large well-selected
samples, that can be compared with other wavelengths in
statistical studies.  With 1~arcsec HPD, confusion will not be a
problem for the most distant (and therefore lower flux) galaxies
(1~nanoCrab). Moreover, these galaxies will have 1-2~arcsec
diameters and so will be easily picked out in deep survey fields.
In 10$^5$s we can get colors for these faint galaxies
($m_B=20-25$). More nearby objects will cover a larger area of
sky, and therefore there will be interlopers.  However, spectra
can identify background QSOs, which can then be used to explore
the cold/warm ISM, via their absorption features.

Broad energy coverage is needed for the study of galaxies, which
encompass soft emission from a hot ISM and stars, `medium range'
emission from SNR and the soft components of some classes of
X-ray binary, and hard emission from the high energy components
of massive binaries and of active nuclei.

\begin{table}
\vspace{-0.3in}
\begin{center}
\caption{Galaxies}
\label{galaxies}
\vspace{0.05in}
\begin{tabular}{|l|l|}
\hline
Distance& With $A$=10~sq.m. and $t$=10$^4$s we can achieve... \\
\hline
Local Group
  &-$R$=1000-10,000 spectra of individual bright sources\\
{}~~to 10~Mpc
  &($L_X>10^{37}$erg~s$^{-1}$ at 10Mpc;$L_X>10^{35}$erg~s$^{-1}$ at M31 \\
  &-$R$=10 (CCD) spectra and colors of fainter sources\\
  &-1~arcsec HPD is needed to resolve sources\\
  &-study uniform samples of galactic sources in different\\
  &~~environments,$\rightarrow$ key factors in source evolution\\
  &$\rightarrow$Study SS433 double-jet sources out to M51\\
  &$\rightarrow$Binary doppler shifts (and so masses) in M31\\
  &$\rightarrow$SNR turbulent velocities in LMC\\
  &$\rightarrow$spatial and spectral properties of the hot ISM\\
\hline
D$\le$Virgo
  &- $R$=1000 spectra of all galaxies ($L_X
>10^{39}erg~s^{-1}$)\\
{}~~($\sim$20~Mpc)
  &-spatially resolved spectra of galaxies  ($L_X >
10^{40}erg~s^{-1}$)\\
  &$\rightarrow$ study temp., density structure, chemical
   composition,\\
  &~~ motions of hot ISM of E and S0 galaxies, measure masses \\
  &$\rightarrow$Supernova expansion velocities \& abundances to Virgo\\
\hline
D$\le$Coma
  &- $R$=1000 spectra of all galaxies ($L_X >10^{40}erg~s^{-1}$) \\
{}~~($\sim$100~Mpc)&(includes bright S, and large fraction of E and S0)\\
\hline
D$\le$500~Mpc
  &- $R$=10 (CCD) spectra of all galaxies, $L_X
>10^{40}erg~s^{-1}$\\
\hline
Z$\le$1
  & bright E galaxies ($L_X > 10^{42}erg~s^{-1}$)\\
  & $\rightarrow$ chemical evolution of galaxy ISM\\
\hline
Z$\le$2
  &-$R$=10 (CCD) spectra of:\\
  & cD galaxies, groups ($L_X > 10^{43}erg~s^{-1}$)\\
  &- all clusters of galaxies\\
  &- rich distant clusters of galaxies\\
  &$\rightarrow$Chemical evolution of the intracluster medium\\
  &$\rightarrow$Mass motions in cooling flows \& clusters\\
\hline
\end{tabular}
\end{center}
\vspace{-0.3in}
\end{table}

\medskip
\noindent\underline{\em Stellar Winds:}
Stellar mass loss is a major contributor to the chemical
evolution of the galaxy, so direct determination of abundances
from X-ray spectra in stellar winds is important. X-rays are
especially good for abundance determinations.  At a spectral
resolution of 1000, one can measure Doppler broadening and line
profiles of X-ray lines emitted in the winds of OB stars (whose
winds are believed to flow at speeds of 1,000~km~sec$^{-1}$);
such studies will thus be definitive for testing models in which
the X-ray emission is thought to come from shocked wind material
embedded in the outflows.  The distribution of plasma at
different temperatures, using spectral emission lines to derive
emission measures, can be tomographically mapped using stellar
rotation.

\medskip
\noindent\underline{\em Solar System:}
Medium resolution spectra can use X-ray fluorescent lines to give
chemical composition and ionization state of the magnetospheric
material of Jupiter and Saturn. Comets were unexpectedly bright
and common in the ROSAT Sky survey. Their chemical composition
can be studied with high resolution X-ray spectroscopy.

\medskip
\noindent\underline{\em All Sky Surveys:}
ROSAT has produced a sky survey with 10$^5$ sources in the soft
X-ray band, and ABRIXAS will soon complement this with similar
numbers in the hard band (Tr\"{u}mper, this workshop). Deeper
surveys are highly desirable. A factor 100 more sources over the
whole sky will complement well the current large area optical
(Sloan DSS) and radio (NVSS, FIRST) surveys.

A 10~m$^2$ telescope could provide its own `Schmidt Survey':
Following the ROSAT model, a 6-month sky survey (for a
30$^{\prime}$~dia. field of view) would reach a limiting flux
density of 0.2$\mu$Crab, equal to that of the ROSAT `Lockman
Hole' Deep Survey (Hasinger et al., 1993). This would produce
some 10 million sources on the sky. This Sky Survey could, for
example detect and resolve normal galaxies well beyond Coma, a
volume 3000 times larger than the RASS; and gather CCD spectra of
rare coronal stellar populations, i.e. objects in the
low-luminosity tail ($\approx$ 10$^{25}$ ergs s$^-1$) of stellar
X-ray luminosity functions for low-mass stars.  The essential
need for high angular resolution X-ray surveys was shown by the
ROSAT HRI detection of large numbers of T Tauri stars missed by
optical surveys. This has important repercussions on our
understanding of the stellar IMF (Feigelson 1996) and so on star
formation theory and galaxy evolution models. The survey would
also provide a definitive study of the change (if any) in stellar
activity levels as stars become fully convective on the main
sequence.

This survey would simultaneously produce a survey in the
2--10~keV band to 1$\mu$Crab, giving $\sim$4 million sources
(assuming the Ginga fluctuations logN-logS normalization).  Being
relatively unaffected by absorption, both Galactic and within
sources, this hard band survey would go deep into the source
population invoked to explain the discrepancy between the
ROSAT and ASCA $logN-logS$.

The Fe-K 6-7~keV band would be covered too, and with CCD (or
better) resolution the Fe-K line could be detected down to
20$\mu$Crab for normal thermal sources. Extraordinary sources
such as NGC~6552 which have 15\% of their flux in the Fe-K line
(Fukazawa et al 1994) would be picked out at 1$\mu$Crab.

Other types of surveys are also matched well to 10~sq.meters:
`Rapid Surveys' of large samples of objects are essential to
building statistical samples, classification of objects, and
discovery of novel phenomena. At 10$\mu$Crab a 10~sq.meter
collecting area will take 1000~s to obtain a high quality CCD
resolution spectrum. With minimised slew times, normal users can
assemble samples of dozens of objects.

\section{X-ray Astronomy Discovery Space }

ROSAT and ASCA have produced a wealth of excellent new results
(see reviews by Keith Mason and Andy Fabian in this workshop).
However, in the excitement over detecting emission lines and
resolving 2/3 of the cosmic X-ray background we, as a community,
often lose sight of the basic weakness of X-ray astronomy
compared with the UV, optical, infrared and radio astronomies
\footnote{Only Gamma-ray astronomy is worse off, but it has just
leapt 2 decades with CGRO, and plans to leap another two with its
next mission.}.
By whichever conventional measure of telescope performance is
chosen (Table \ref{comparison}) X-ray astronomy is about a factor
of 100 behind: angular resolution, spectral resolution,
collecting area. Worse, to a great degree the other astronomies
have these better capabilities all at once, whereas our
telescopes specialize in one or another.

\begin{table}
\vspace{-0.3in}
\begin{center}
\caption{X-ray compared with other Astronomies}
\label{comparison}
\vspace{0.05in}
\begin{tabular}{|l|l|}
\hline
X-ray & UV/optical/IR/radio\\
\hline
5 arcsec & 0.1$-$1 arcsec\\
E/$\Delta$E =1-50 & E/$\Delta$E = 1000$-$10,000 \\
A$_{eff}$=0.02 m$^2$&A$_{eff}$= 3$-$75 m$^2$ \\
N$_{sources}$=10$^5$&N$_{sources}$=10$^8$ \\
\hline
\end{tabular}
\end{center}
\vspace{-0.3in}
\end{table}

Being far behind the other astronomies would not matter so much
if X-ray astronomy were catching up. There was a long period,
from the birth of cosmic X-ray astronomy to the end of the {\em
Einstein Observatory}, when X-ray astronomy advanced in
sensitivity by a factor of 100 every 10 years, 2 decades per
decade (Table \ref{progress})- from sounding rockets to
satellites to grazing incidence imaging. In those heady years
being the underdog coming from behind was exciting. We were
catching up the optical astronomers. We felt that soon we would
be true equals. It didn't happen.

\begin{table}
\vspace{-0.3in}
\begin{center}
\caption{X-ray Astronomy's Rate of Progress}
\label{progress}
\vspace{0.05in}
\begin{tabular}{|l|l|l|l|l|l|}
\hline
Year& Detection & Factor & Spectrum & Factor  & Missions \\
    & Threshold &Increase& Threshold&Increase&        \\
    &  ($\mu$Crab)  &        &  ($\mu$Crab)&  &        \\
\hline
1966&100000     &---    &$\sim$10,000,000& ---    &rockets\\
    &           &       &(solar flare)&       &           \\
1976&1000       &100     &1,000,000  &10      &{\em Uhuru},Ariel~V \\
    &           &       &(Bragg)     &        &SAS-3      \\
1986& 10        &100    &10,000      &100     &{\em Einstein}\\
    &           &       &(grating)   &        &           \\
1996& 1         &10     &1000        &10      &ROSAT, ASCA\\
    &           &       &            &        &           \\
2006& 0.1       &10     &100 (CCD)   &10      &AXAF, XMM  \\
    &           &       &1000 (gratings,&10      &ASTRO-E \\
    &           &       &calorimeters)&       &           \\
2016& 0.001     &100    &10          &100     &10~sq.meter\\
    &           &       &            &        &           \\
\hline
\end{tabular}
\end{center}
\vspace{-0.3in}
\end{table}

Had AXAF been launched on schedule that pace would have been
maintained. Political problems, budgetary restrictions and tragic
accident all contributed to the delay of AXAF. But whatever the
causes, the effect has been to reduce the pace of X-ray
astronomy's progress to one decade per decade. While X-ray
astronomy was on hold, the optical, UV and IR astronomies have had
HST, Keck and IR-arrays come into being to revolutionize
them. X-ray astronomy now is no longer catching up, it is merely
not slipping behind.


While important science drivers (as in \S 2) must be presented
to justify any new major scientific project, the real advantage
of a factor of 100 leap are the discoveries that cannot be
foreseen. When making leaps of sensitivity by two orders of
magnitude or more the only type of argument that works is the
``Discovery Space'' (Harwit 1984) kind. We ask ``In order to go a
factor of 100 fainter and so open up new volumes of discovery
space, what do we need for: area (to get enough counts)?; angular
resolution (to cut down background, and to locate sources)? How
many counts do we need to accumulate a good spectrum? This style
of argument is highly robust against changes in models and even
to new discoveries. We use it in this paper.

The May 1980 proposal to NASA for a Large Orbiting X-ray
Telescope (`LOXT')
\footnote{The Large Orbiting X-ray Telescope (`LOXT') was first
proposed in May 1970 (Tucker \& Giacconi 1985), before the launch
of $UHURU$ (December 1970). Only 20 cosmic X-ray sources were
then reliably known (Seward, February 1970), all based on
$\sim$5~minute sub-orbital flights, most with 1~degree-like
angular resolution. The faintest source was about 40~mCrab. In
essence it combined AXAF and XMM, strapped together and
co-pointing on a single satellite (a 1.2$^m$ dia., 2~arcsec
mirror; plus a 0.5m$^2$, 10~arcsec mirror).  At the time of the
LOXT proposal (which was accepted by NASA) the construction of
such telescopes was well beyond the level of demonstrated
technology. Skylab would not fly for 3 more years, and it would
have 0.3$^m$ dia. and $\sim$20~arcsec imaging (see e.g. Belew and
Stuhlinger 1973). Imaging X-ray detectors had not been
developed. (Skylab used film). X-ray microchannel plate imagers
were only a concept. However, the goal of a large grazing
incidence imaging telescope provided the scientific motivation
for both an optics and a detector program. The funding agency
could see the overall goal spelled out, so it could appreciate
why the scientists needed the technology development money.  }
provides an excellent example of the kind of justification to use
for `factor of 100' missions. Justified on the basis of discovery
space arguments the LOXT program resulted in a strong NASA X-ray
optics and detector R\& D program.  A more recent example is that
of the Next Generation Space Telescope `NGST'
\footnote{
The report `HST and Beyond' (1996, ed. A. Dressler) sets a bold
goal to succeed HST based solely on science considerations: the
construction of a 6-8~meter diameter infrared optimized
telescope, diffraction limited at 2~$\mu m$ (0.2~arcsec), with
its primary mirror cooled to 30-50~K, preferably in Jupiter
orbit, to reduce the zodiacal light background (the dust that
emits the zodiacal light is cooler there).  These goals are
technologically ambitious. No-one knows how to carry them out,
especially within a plausible budget. Yet this report has led
this year to a new NASA program of several million dollars per
year, directed at carrying out its prime recommendation.}.
With a long term goal in place the agency can understand why
particular technologies need to be developed.

We must do the same in X-ray astronomy. X-ray astronomy goals
must be bold and science driven. Our goals must capture the
imagination of the astronomy community, of the funding agencies
and the public. This should not be hard. X-ray astrophysics
encompasses much of the galaxy evolution and galaxy ecology
targeted by NGST, and adds the twin theme of exotic physics
through black holes, neutron stars, supernovae and quasars.

To appeal to a broad astrophysics community a mission must be
sensitive enough to reach the whole enormous range of objects
that we know emit X-rays, not just specialist High Energy
objects. It also means that it must make good observations of
these objects in a few hours, else too few observations will be
taken to sample the richness of the X-ray sky.

To accomplish this we must take orders of magnitude steps and
open new discovery space. We must accelerate X-ray astronomy
toward the capabilities of optical, UV and radio astronomy.

\section{Deriving the Telescope Goal}

\subsection{A 10 Square Meter Collecting Area}

We begin with the concept of the `baseline
observation'. Everything then scales from this. The baseline
observation can be defined without reference to any specific
target. It relies solely on discovery space arguments.  The
starting point is to ask what a {\em typical} observation should
be. This has four parts: What flux source ($f$)? What exposure
time ($t$)? What detection efficiency ($e$)? How many counts
($N$)? Putting these together defines the collecting area ($A$):
$$ N~ =~ f~ .~ A~ .~ t~ .~ e ~/~<h\nu >$$
$f$ is determined by the type and number of sources one must
reach; $t$ is determined by how many observations/year one should
carry out; and $N$ is determined by the type of measurement one
wants to make. ($<h\nu >$ is the mean photon energy.) Even if you
disagree with our choice of $fAte$, this approach allows us to
compare proposed missions quantitatively.

With this approach we derive a baseline area $A$=10~sq.m. This is
derived from a baseline $t$=10$^4$s, $N$=10$^5$ counts,
$f$=10$\mu$Crab, and $e$=1. A 10$\mu Crab$ source gives about 1
count/s/square meter (in the ROSAT band). (For grating
spectroscopy $e=1/4$ applies and hence longer $t$). We assumed
$<h\nu >$=1~keV.  These baseline values are justified below.

\medskip
\noindent{\bf Flux ($f$) baseline: 10$\mu$Crab}
The typical observation will be of a source at a flux level that
allows a wide choice of object. This must be faint enough that
all classes of X-ray emitter are accessible. We take the ROSAT
All Sky Survey (RASS) flux limit of 10~$\mu$Crab since
many classes of source appear in quantity near this flux.
Examples include: high z quasars, comets (Dennerl et al., 1996),
Shapley-Ames galaxies

\medskip
\noindent{\bf Exposure ($t$) baseline: 10$^4$~s}
To have a healthy scientific discipline observers need to compete
with one another; they need to follow-up on observations that are
puzzling; or on observations that show promise. To support this
the telescope needs to be able to observe a reasonable sized
sample of each type of X-ray emitting object each year.
Let us say there are 100 categories
\footnote{This is clearly a somewhat arbitrary division, but the
number is substantial. The old {\em Einstein Observatory} Catalog
of observations (the ``Yellow Book'', Seward \& Martenis 1986)
already contained about 50 categories. Since then the number has
increased. `Galaxies' for example were not separated into
Ellipticals and Spirals, let alone `hot gas dominated
Ellipticals' vs. `binary dominated Ellipticals'; `Star formation'
was only one category; comets were not mentioned; and so on.}.
Then to make 20 observations in each category requires 2000
observations per year. There are 30 million seconds in a year. At
2/3 efficiency (a plausible, and simplifying value) an average
observation has to last 10,000~s, i.e. 3~hours. That this is
reasonable for serving a large community can be judged by a
comparison with optical usage.  Three hours would be considered a
normal, somewhat long, exposure on a 4-meter class general user
optical telescope
\footnote{This also accords with the experience on ROSAT and
ASCA. ROSAT made $\sim$1500 pointed observations per year
with an average exposure of $\sim$10~ksec. ROSAT publications are
running at 450 publications per year (Tr\"{u}mper, this
workshop). The range of subjects addressed by PSPC observations
spans all of X-ray astronomy (see W\"{u}rzburg proceedings,
Zimmermann, Tr\"{u}mper \& Yorke, eds. 1996), and was used by
many astronomers outside the main X-ray astronomy groups. ASCA
instead observes about 250 targets per year with an average
exposure of around 35~ksec, and publications are of order 60 per
year. ASCA spans a more restricted range of topics (see Makino \&
Tanaka, eds. 1994) and, although in those fields its
contributions are of great importance, ASCA is used
overwhelmingly by `hard core' X-ray astronomers from a small
number of groups.}.

\medskip
\noindent{\bf Counts ($N$) baseline: 10$^5$}
The number of counts we need in an observation depends on the
type of data:

\medskip
\noindent\underline{1. Spectroscopy}
The case for high resolution spectroscopy is well known in X-ray
astronomy. The soft X-ray band contains a high density of atomic
transitions that carry powerful diagnostic information about the
plasma that creates them. Only spectroscopy can extract this
information.

There is, though, no consensus on the resolution $R$
needed. Typical values of 200-300 are usually talked of
\footnote{A physical argument for the $R$ required to remove
blending can be based on the separation of the triplet and
satellite lines for a given ion, indicating $R$=300 to be
sufficient (Holt, these proceedings); this happens to be the
thermodynamic limit of microcalorimeters. Real plasmas, however,
contain multiple ions of each element and multiple
elements. Moreover, astrophysics plasmas rarely have a single
temperature, or density. Still worse, photoionized plasmas are
common. These, in de-exciting from the high levels at which they
recombine, find many more pathways to the ground state, and so
produce quite different lines (Liedahl et al., 1990). In real
astrophysical sources then, far more lines will be present than
the `triplet/satellite' argument suggests. Solar spectra at
$R$=400 are heavily blended (McKenzie et al., 1980) and require
at least R=1000 before the number of emission lines per
resolution element drops to a few.}.
However solar spectra show that $R$=1000 is needed to remove
`line blending' confusion.  and so have unambiguous temperature
and density diagnostics.

The workhorse spectroscopic observation is likely to be an
exploratory spectrum at $R=1000$ \footnote{About the resolution
obtained in routine optical identification programs for X-ray
sources.} (Table \ref{R1000}).
To translate a resolution into a number of counts we need
signal-to-noise criteria. Emission line dominated spectra need
about 20 counts/pixel on average. A bright line (20$\times$mean)
will then be measured to about 5\% , and a continuum point will
be just detected with about 10 counts. Absorption line spectra
need more counts, of order 100 counts/pixel, since the continuum
must be accurately measured in each pixel so that narrow
absorption lines can be individually significant. For $R$=1000
these criteria require $\sim 10^5$ and $\sim 10^6$ counts.

\medskip
\noindent\underline{2. High Resolution Spectroscopy}
At this resolution new physics possibilities open up, such as
Doppler motions and thermal widths (Table \ref{R10000}). For
$R$=10,000 we need $\sim 10^6$ counts for emission line spectra.

\medskip
\noindent\underline{3. Imaging spectroscopy}
This is demanding. A moderate quality CCD spectrum requires 1000
counts, so at 100eV resolution 10$^5$ counts allows us to extract
100 independent spectra from an image. This is quite limited, as
can be seen if you imagine a 10$\times$10 image of a face. At
10eV calorimeter resolution only 10 spectra can be filled.  (Note
that this argument is independent of how the spatial bins are
chosen.)  While line-dominated sources will fare better in the
strong lines, 10$^5$ counts forms a minimal requirement.

\medskip
\noindent\underline{4. Imaging: Dynamic Range}{\em (brightest
peak/faintest detected features)}
Dynamic range determines image quality.  Astronomy, being a
logarithmic subject, needs large dynamic range. Faint features
often reveal the physics. For example, the powering of radio
galaxy lobes was puzzling before the VLA imaged the narrow plasma
jets directed from the active nucleus to the lobes, many kpc
distant. The VLA detected these jets not because of improved
spatial resolution, but because of its much better dynamic range
of $\sim$1000.

By comparison the best X-ray images from ROSAT have tiny dynamic
range. The ROSAT Calendar picture of complex structure in the
NGC~1275 cooling flow (Sept 1994, also B\"{o}hringer et al.,
1994) has a dynamic range of only a few. (Excluding the nuclear
point source.)  If the faintest pixels have marginal, 10 photon,
detections, then a dynamic range of 1000 requires 10$^4$ photons
in the peak pixel. So 10$^5$ photons are needed in the source.

\subsection{Angular Resolution: 1 arcsec. HPD}

The second key parameter in any telescope design is angular
resolution. The primary reason that we want high angular
resolution is that astrophysical sources show structure on a the
finest scale we have yet imaged. But this is a squishy standard.
Fortunately there are some quantifiable criteria and these lead
to quite high angular resolution requirements, half power
diameters (HPD) of a few arcseconds
\footnote{
Proposals always quote a resolution and an effective area, but
these two numbers are linked. For example when using a Half Power
Diameter for spectroscopy, the effective area should be divided
by two since half the counts (by definition) are outside the HPD.
A more appropriate measure of resolution for spectroscopy would
be the 80\% PD, closer to the typical extraction region used in
practice. But then the angular resolution of the mirror will be
2.5-5 times worse (using XMM as an example, Willingale, this
workshop). We will scale all results to the HPD.}.
We propose the angular resolution goal of 1 arcsecond HPD.

\medskip
\noindent\underline{1. Detection Confusion}
The best known criterion is `source confusion'.  To avoid this
problem radio astronomers developed a flux limit criterion for
telescopes (Murdoch, Crawford \& Jauncey, 1973). Essentially this
is a Poisson statistics problem.  If we adopt the criterion that
{\em ``A detection should due to only one source at the
3.5$\sigma$ gaussian confidence level''} then the mean density of
sources on the sky comes out to be 1 per 40 beams.

How this translates to a beamsize depends on knowing the
logN-logS relation for the sources in question. From ROSAT we now
have a good handle on this down to faint X-ray fluxes. Applying
the 1 in 40 criterion then at the 1$\mu Crab$ level (`ROSAT
Medium Survey' level) we need a half power diameter (HPD) of
14~arcsec. For the hard 2-10~keV band, the higher logN-logS
normalization (Inoue et al., 1996) brings this down to 8~arcsec.
At the ROSAT Deep Survey limit (0.1$\mu Crab$) the HPDs come down
to 4~arcsec and 2~arcsec respectively. At faint fluxes
(0.1$\mu$Crab) extragalactic sources are not randomly
distributed, but follow the large scale structure distributions
of galaxies (Refregier et al., 1996), implying HPD$<$4~arcsec.

\medskip
\noindent\underline{2. Spectral Confusion}
There is another type of confusion that is important even for
10$\mu$Crab (RASS limit) sources. If we saw an unexpected
emission line at a strange energy in a source spectrum, we would
need to have confidence that this was not due to a background or
foreground contaminating source
\footnote{The most famous recent example of this class of problem
was the report of periodic variations in the Seyfert galaxy
NGC~6814 (see Mittaz \& Branduardi-Raymont, 1989, Done et
al,. 1992). The short, stable period implied an accurate
clock, ruling out all starburst AGN models, and strongly
implying a black hole.  Although only one AGN was involved
it mattered hugely and for 5 years many papers were written
chewing over the data and its interpretation. All for
nought. In all observations prior to ROSAT another bright
X-ray source, a galactic binary of the cataclysmic variable
class, had been unknowingly included in the beam and had
produced the periodicity (Madejski et al., 1994). ({\em
Einstein} and EXOSAT-LE missed the CV because their
field-of-view was too small.) NGC~6814 was a member of a
sample of the Piccinotti et al., (1982) sample which was
not source confused on the standard criterion by a factor
of $\sim$5.}.
. Highly obscured AGN emit a reflection spectrum which is
dominated by large equivalent width emission lines
(e.g. the Circinus galaxy, Matt et al. 1996). Even if such
a source were 20 times fainter than the target it would
still produce strong lines (up to EW$\sim$100 eV). Such
objects are likely major contributors to the population
that produces the enhanced logN-logS above 2~keV, so they
may be common. Foreground M stars (particularly during
flares) are another source of spectral confusion. (Brief
observations with higher resolution instruments won't
detect the lines, so won't solve the problem.)

A strict criterion is needed when looking for peculiar
features in individual objects. If we adopt the criterion
that {\em ``A detection should due to only one source at
the 5$\sigma$ gaussian confidence level''} then the mean
density of sources on the sky comes out to be 1 per 1000
beams. Because of line dominated sources a source density
of 200/sq.degree (i.e. for sources at least 20 times
fainter than the baseline, 10$\mu Crab$, target must be
used, as well as an 80\%HPD extraction region.
\footnote{
Proposals always quote a resolution and an effective area,
but these two numbers are linked. For example when using a
Half Power Diameter for spectroscopy, the effective area
should be divided by two since half the counts (by
definition) are outside the HPD.  A more appropriate
measure of resolution for spectroscopy would be the 80\%
PD, closer to the typical extraction region used in
practice. But then the angular resolution of the mirror
will be 2.5-5 times worse (using XMM as an example,
Willingale, this workshop). We will scale all results to
the HPD.}.
This leads us to a HPD$\leq$3~arcsec ($\leq$2~arcsec above 2~keV).

\medskip
\noindent\underline{3. Source Identification}
At 1$\mu Crab$ ROSAT PSPC error circles (about 25~arcsec
dia.)  contain 2-3 optical candidates at 22$^m$ (Jones et
al., 1996). To identify the final 1/3 of the X-ray
background we need to $\sim$1~nanoCrab, and so $m_r
\sim$28. At this level there is 1 galaxy per 4~arcsec
dia. beam (Tyson, 1988). This then requires source
centroiding to 2.5~arcsec to get a few percent chance
co-incidence rate. Since the faintest sources are always
$\sim$3$\sigma$ detections this implies a HPD$<$4~arcsec.

\medskip
\noindent\underline{4. Background Reduction}
To reduce the background to 10\% of the source counts at
10$\mu Crab$ (RASS limit) level requires a $\sim$80\%PD
source extraction region
an HPD$<$20~arcsec.

\medskip
\noindent\underline{5. Rich Fields}
There are many places where sources are not randomly
distributed.  Clusters of point sources are found in star
formation regions, spiral galaxies, HII regions, starburst
galaxies, globular clusters, and individual galaxies in
clusters of galaxies (detectable to Coma with a
10~sq.~m. telescope). In each case the HPD is reduced by
(source density)$^{1/2}$, i.e. HPD$<$4~arcsec.

\medskip
\noindent\underline{6. The Real Reason: Complex Structures}
The ROSAT HRI has shown us complex structures down to its
5~arcsec HPD in: Clusters of galaxies, active galaxies,
elliptical and spiral galaxies, supernova remnants,
globular clusters, star formation regions, and other
places.  (see W\"{u}rzburg meeting proceedings,
eds. Zimmermann, Tr\"{u}mper \& Yorke 1996).

There is every reason to believe that we will continue to
see complex structure with the 1~arcsec HPD of AXAF. We
already know that UV, optical, near-IR and radio emitting
objects have complex structures at the 1~arcsec level. The
very same objects are X-ray sources. It is hard to imagine
a general argument that says structure will be smooth in
the X-ray band when it is complex at all longer
wavelengths. Observing an inhomogeneous object with poor
resolution can reveal averaged properties, but this is
qualitatively inferior to seeing directly which temperature
lies where and how they are juxtaposed. The minimum angular
resolution goal should be 1~arcsec
\footnote{
We should also not ignore the possibility of finer
resolution.  VLBI shows that milli-arcsecond structures can
be seen in X-ray emitting objects, and micro-arcsecond
resolution should resove binaries and AGN (Cash, this
workshop). A long term investment into the feasibility of
X-ray interferometry should be part of our program.}.

\subsection{Field of View: $>$10~arcmin dia.}

Since Wolter-like optics have a $\sim$1/2~degree diameter
field of view with good imaging, it is wasteful not to use
it. Large fields of view enable: (1) serendipitous
discoveries; (2) large area, even All-Sky, surveys; (3)
accurate background subtraction; (4) large object imaging
(Abell clusters, Shapley-Ames galaxies, star formation
regions, and many others).

If all large objects and associations of objects had
already been studied extensively, then a narrow
field-of-view mission may make sense if, for example, a
major mission simplification resulted. However, in X-ray
astronomy we have not studied even the nearby objects yet
in any detail (e.g. NGC~1275, above).  Based on the sizes
of nearby galaxies and clusters of galaxies, and the
expected surface density of the faintest sources a field of
view of at least 10~arcmin. is needed
\footnote{
A small field of view can be fine in some circumstances
(e.g. WFPC-2 on HST).  But even with a narrow field of view
many independent pixels are usually needed. The trend in
ground-based telescopes instead is to create a multiplex
advantage by building large field of view optics that can
study hundreds of objects at once.}.

\section{Instrumentation Goals}

We have derived from fundamental considerations that to
enter the realm of capability enjoyed by the other
astronomies, X-ray astronomy needs a mirror collecting area
of 10~sq.meters focusing to 1~arcsecond. Spectroscopy at
$R$=1000-10,000 is needed to exploit the richness of X-ray
atomic physics.

There are some general points about the hardware for such a
mission we need to make: (1) A major feature of a large
area, good angular resolution X-ray telescope is the large
variety of instruments to which it could fruitfully feed
photons (see Table \ref{instruments}). Clearly large field
imagers and high resolution ($R>$1000) spectrometers are
both essential.  (2) Grating spectroscopy is a major
priority (see next point). (3) All instruments must have
short (1~msec) readout times (4) The energy range for which
the telescope is optimized should emphasize the lowest
energy X-ray band (see below). (5) we should consider
designs that allow an X-ray telescope to be upgraded with
state-of-the-art instrumentation.

\small
\begin{table}
\vspace{-0.3in}
\begin{center}
\caption{Instrumentation for the 10 sq.meter X-ray Telescope}
\label{instruments}
\vspace{0.05in}
\begin{tabular}{|p{5in}|}
\hline
{\bf Focal reducers/Re-imagers}
Large collecting area provides a tolerance for the loss
of light extra reflections produce. Additional optical surfaces
allow the whole gamut of traditional optics devices to be used
for X-ray astronomy: collimators, re-imagers, focal reducers.
These will have wide application in feeding new instruments.
\\ \hline
{\bf Wide Field Imagers}
Wide field imagers are essential to surveys, and the study of
extended objects. CCD spectral resolution is all that can be
usefully employed to fill arrays (see Table 5, \S 5.3).
Larger $R$, e.g. with cryogenic devices, is of course desirable.
Re-imaging to a plate-scale optimized for existing detectors may
be desirable.
\\ \hline
{\bf $R$=1000 Spectrographs}
Basic X-ray plasma diagnostics become feasible only at $R$=1000
(see \S5.3).
\\ \hline
{\bf $R$=10,000 Spectrographs}
New physics (Doppler motions, thermal line widths) become
accessible at $R$=10,000
(see \S5.3).
\\ \hline
{\bf Slit Spectrographs}
Current X-ray spectrometers are slitless. This leads to confusion
and loss of resolution when observing crowded fields or extended
sources. Many strong line emitting sources are extended (SNR,
star forming regions, clusters of galaxies, galaxies) With
collimating and re-imaging optics a true X-ray slit spectrograph
could be built.
\\ \hline
{\bf Spectral Re-imagers}
Komykhov lenses act like optical fibers in X-rays and are used in
medical applications They are likely to be inefficient, but with
enough photons they could be usefully employed to create a
spectral-spatial data cube. A bundle of lenses could sample a 2-D
image and re-image it into a linear array. This array could then
be passed through a grating to form high resolution spectra.
\\ \hline
{\bf Polarimeters}
Inherently hard since a 10$\pm 3$\% measurement requires the
count rate measured to $\pm$0.3\%, i.e. 10$^6$ counts. For 10\%
efficiency of polarization detection ($e$=0.1) this measurement
can be carried out on a 100~$\mu$Crab source in 10~ksec.
\\ \hline
\end{tabular}
\end{center}
\vspace{-0.3in}
\end{table}
\normalsize

\subsection{Grating Spectroscopy}

High resolution soft X-ray spectroscopy can only be carried
out with dispersive techniques. Thus the current enthusiasm
for cryogenic non-dispersive spectroscopy (see this
workshop) needs some moderating.  Microcalorimeters have
good points, in particular the clear factor of 3-4 QE gain
over dispersive spectroscopic systems, and the prospect of
an `ideal' imaging spectroscopy detector. However they also
have limitations:

(1) The resolution of a few hundred now in prospect from
microcalorimeters will not be enough.  Current
microcalorimeters have a thermodynamic limit of 1-2~eV or
about $R$=500 at 1~keV, and other noise sources may cut in
before that limit is reached. The current best laboratory
resolution is $\sim$5~eV, i.e. $R$=200 at 1~keV. To do
better with non-dispersive methods does not yet seem
feasible. (2) They have limited ($\sim 10\times 10$) array
size. Larger arrays require a long and expensive research
and development program and may not be feasible (Cooper,
this workshop). (3) They require complex blocking filters
that limit low energy response. (4) They need ultra-low
temperature cryogenics, which constrain satellite orbit
(away from LEO), instrument lifetime, and increase
instrument complexity and mass. Of course a new, perhaps
even colder, class of non-dispersive instrument may come
along.

Dispersive techniques are thus essential. This primarily
means diffraction gratings. Dispersive grating spectroscopy
is essential both for E$<$2~keV and for $R\sim$10,000
spectroscopy.  Fortunately, the price of going to
dispersive spectroscopy is not excessive: a factor of 3-4
in efficiency ($e$), plus more complex optics to enable
slit spectroscopy. Development of grating optics,
especially for slit spectroscopy, must be a high priority.

\subsection{The Softest Energy Band: 0.1-0.5~keV}

The improvement in microcalorimeter resolution at higher
energies is helping to drive the X-ray astronomy
communities interests away from the sub-1~keV band. Much
effort is going into extending the response of grazing
incidence X-ray optics to the higher energy 10-50~keV band
(e.g. Elvis, Fabricant and Gorenstein 1988, Gorenstein,
this workshop).

The soft, 0.1--0.5~keV band has the same 0.7 decade
bandwidth.  as 10-50~keV and is scientifically far more
important: (1) the soft band contains 10$^4$ times the
number of photons (for a Crab spectrum), albeit dependent
on Galactic absorption; (2) the soft band contains a huge
number of atomic transitions as evidenced by solar spectra
and EUVE; (3) as we observe to higher redshifts the
familiar 0.5-2~keV lines and edges move into this band. By
z=4, where many quasars are already known, the soft band
has become 0.5-2~keV in the rest frame. However, a 2~eV
microcalorimeter is reduced to $R$=100 at 0.2~keV. A
grating spectrograph gains resolution at low energies and,
using unfiltered detectors could be designed to work
efficiently at $R>$1000 down to 0.1~keV.

\section{What did the Workshop Show?}

The challenge of the goals set out above attracted many
comments during the NGXO workshop. Somewhat to our surprise
the goals do not look far-fetched. Optics technologies were
described that approach (even exceeding) the goals.

Of the optics technologies discussed, microchannel plate
optics are 100 times ligher per unit collecting area than
replica optics and so will be valuable in large area `light
bucket' missions. However they seem unlikely to reach
1~arcsecond HPD (Fraser, this workshop) and so are not a
candidate for the 10sq.meter telescope. Foil optics seem
unlikely to approach the necessary angular resolution
(Serlemitsos, this workshop) and are 10 times heavier than
microchannel plates per unit collecting area. Replica
optics have a better chance to achieve one arcsecond
(Willingale, this workshop).

More ambitious optics, such as the spherical multi-element
systems described by Cash (these proceedings) or adaptation
of the large optical telescope methodologies (e.g. Angel,
private communication) have been little explored. These
need development funding. Cash's spherical optics (Table
\ref{optics}, and this workshop) suggest that it may now be
feasible to reach 0.1~arcsec imaging performance in X-ray
astronomy.

\small
\begin{table}
\begin{center}
\caption{X-ray Optics Technologies}
\label{optics}
\vspace{0.05in}
\begin{tabular}{|p{5in}|}
\hline
{\bf Replica optics}
developed from XMM can be expected to reach HPD=5~arcsec,
possibly 3~arcsec. With existing shell thicknesses Ariane~V can
lift 2-3~sq.~m. into orbit (Willingale). Further lightweighting
techniques might increase this by a factor of 2-3 (Citterio,
Tananbaum). The angular resolution is still likely to be
inadequate however.
\\
\hline
{\bf Spherical optics}
seem to have major potential. To achieve $<$0.25~arcsec HPD at
8~keV on a first attempt is impressive. The low cost per unit
collecting area is excellent. The error analysis showing that
1/100~arcsec HPD is plausible makes further development of these
optics a top priority. If the sharp reduction in cost per
kilogram to orbit being pursued by NASA comes to pass by 2005 as
planned, then concerns about mirror weight will become less
central. However, thinned optics designs must be worked on to get
10~sq.meters.
\\
\hline
{\bf Large optical mirror technology} has developed greatly in
the last decade. Large light-weighted thin mirrors with good
surface roughness have been built including a 2mm thick, 1m
dia. prototype for NGST. Many of the techniques used at the
Univ. of Arizona Mirror Lab., for example, could be adapted to
the grazing incidence case (Angel R., private communication.)
\\
\hline
\end{tabular}
\end{center}
\end{table}
\normalsize

A major weakness of the workshop was the lack of
consideration of high spectral resolution using grating
spectroscopy. Since this is the only route to the necessary
goal of $R>$1000 spectroscopy, so we must pursue it.

Three main missions emerged from the workshop as the next
step.  However, they only partially address our
goals. These missions overlap considerably in their aims
and concept: (1) a European follow-on to XMM that takes
replica mirror technology another step ($\sim$ 1-3m$^2$;
$\leq$5'' HPD); (2) a Japanese successor to ASTRO-E
($\sim$1 m$^2$, $\sim$30''[?]HPD) using improved foil
optics, and launched with the new large H-2; and (3) the US
`High Throughput X-ray Spectroscopy' mission (HTXS) ($\sim$
1m$^2$; $\leq$15'' HPD).

In their collecting area these plans are only a factor of
5-10 too small. In angular resolution however they all are
inadequate. Since all three missions stressed calorimeters
and 10--50~keV response none reaches the spectral
resolution and low energy coverage that will turn X-ray
astronomy into X-ray astrophysics.

\section{What is to be done?}

How can X-ray astronomy reach the 10~sq.m/1~arcsec goal?
As a community we need to become less `political' and more
science driven. Presently we act tactically, reacting to
each announcement of a mission opportunity by putting
together good proposals. Since a good proposal requires the
use of existing, proven technologies, we do not get beyond
incremental advances. Instead we must get strategic. We
must set long term science goals, as we have done here, and
then {\em begin!} Start work now on achieving our goals
through technology studies.

The primary thrust of these technology studies must be
innovative X-ray optics: the pursuit of {\em several} new
mirror technologies. These studies need substantial
funding, of order \$ 2M/year each (i.e. 5 man-years/year,
including overhead, plus equipment). A \$ 10M/year program
sounds huge compared with the less than \$ 0.5M/year now
spent by NASA on X-ray optics, outside of flight
programs. It is small potatoes, though compared with the
astrophysics budget at NASA. It is similar to the NASA
investment in ASTRO-E or a SMEX, yet the pay-off is hugely
greater.

A second but still very important study must address the
spacecraft and space infrastructure needed to support a
10~sq.m. high resolution telescope. Areas of importance
include: long, low weight structures; active control of
optical benches; moment-of-inertia balancing for fast
slewing; servicing from Station, Shuttle and small
launchers. Topical technical workshops on each of these
areas would be a good way to focus the issues and technical
challenges and to identify the most promising areas of
work.

The US, European and Japanese intermediate missions presented at
the workshop should not take all the energies of the community
and so prevent a strong development program. The goal we
originally outlined seemed bold. This workshop has shown that in
fact it is quite plausible, and only a stretch from plans now
being formulated.

To reach 10~sq.~meters in 10 years (5 years of development and 5
years to build) brings us back on track to advancing the field by
a factor of 100 per decade.  Can we really do this? Only if we
try.

\bigskip
\bigskip
\centerline{\bf Acknowledgements}

We wish to thank many of our colleagues for stimulating
discussions, especially Andrew Szentgyorgyi, Nancy Brickhouse,
John Raymond, Richard Willingale, Ronald Polidan, Robert Rosner,
Webster Cash, Roger Angel, Andrew Fabian, John Gibbons, Steven
Murray, Harvey Tananbaum, Salvo Sciortino and Giorgio Palumbo.

\newpage

\end{document}